\documentclass[12pt]{iopart}

\usepackage{epsfig}
\usepackage{makeidx}
\usepackage{graphicx}
\usepackage{epstopdf}
\usepackage{subfigure}

\usepackage{amssymb}

\begin{document}

\title[Economic inequality and mobility for stochastic models]{Economic inequality and mobility
for stochastic models with multiplicative noise\ \footnote{\ This research was 
funded in part by the Free University of Bozen-Bolzano through the project DEFENSSE}}

\author{Maria Letizia Bertotti $^{1}$, 
              Amit K. Chattopadhyay $^{2}$, 
              Giovanni Modanese $^{1}$
\\
%
%
}

\address{
$^{\,1}$ Free University of Bozen-Bolzano, 
Faculty of Science and Technology, \\
39100 Bolzano, I \\
$^{\, 2}$ Aston University, 
Engineering and Applied Science, \\
Birmingham, B4 7ET, UK
}

\ead{marialetizia.bertotti@unibz.it, a.k.chattopadhyay@aston.ac.uk, giovanni.modanese@unibz.it}
\begin{abstract}
In this article, we discuss a dynamical stochastic model that represents the time evolution of income distribution of a population, where the dynamics develop from an interplay of multiple economic exchanges in the presence of multiplicative noise. The model remit stretches beyond the conventional framework of a Langevin-type kinetic equation in that our model dynamics is self-consistently constrained by dynamical conservation laws emerging from population and wealth conservation.
This model is numerically solved and analyzed to interpret the inequality of income as a function of relevant dynamical parameters like the {\it mobility} $M$ and the {\it total income} $\mu$. In our model, inequality is quantified by the {\it Gini index} $G$.
In particular, correlations between any two of 
the mobility index $M$ and/or the total income $\mu$ with the Gini index $G$ are investigated and compared with the analogous correlations resulting from an equivalent additive noise model. Our findings highlight the importance of a multiplicative noise based economic modeling structure in the analysis of inequality. The model also depicts the nature of correlation between mobility and total income of a population from the perspective of inequality measure.
\end{abstract}

\maketitle

\section{Introduction}

Various  approaches inspired by a combination of statistical physics and kinetic theory have been proposed in recent years 
for the description of economic exchanges and market societies, see for example \cite{ChaCha,DurMatTos,SinCha,YakBar,PatHeiCha,BM1,PatCha}.
In these approaches, individuals trading with each other are identified as particles or gas molecules 
which undergo collisions. Methods and tools based on physics have proved useful 
also in such socio-economic contexts to investigate the emergence of macroscopic features
from a whole of microscopic interactions.
With this perspective, some mathematically founded market economy models,
characterized by the ability to also incorporate  taxation and redistribution processes,
have been proposed and studied in \cite{BM2,BMonAMC}. In these papers, society is equated to a system composed by 
a large number of heterogeneous individuals who exchange money through binary and other nonlinear interactions 
and are divided into a finite number $n$ of income classes. 
The models are expressed by a system of $n$ nonlinear ordinary differential equations of the kinetic-discretized Boltzmann type,
involving transition probabilities
relative to the jumps of individuals from a class to another. 
The specification of these probabilities and of the parameters
which define the trading rules, including the tax rates pertaining to different income classes and other properties of the system, determines the dynamics.
Collective features like the income profile and related indicators like the Gini index - here, a measure of economic inequality - 
result from the interplay of a range of such interactions. 
Due to the presence of the mentioned transition probabilities, the process is stochastic \cite{AY} but the differential equations 
governing the evolution of the fraction of individuals in the classes are deterministic.

In real world, however, the time evolution of an economic system is governed not only by fixed rules and parameters:
it is subject to the effects of unpredictable perturbing factors as well.
To consider the influence of these factors, we recently introduced a 
Langevin-type  kinetic model \cite{BCM}, incorporating an Ito-type additive noise term into the set of dynamical equations. 
Several numerical simulations provided evidence of the persistence of patterns 
already established in the deterministic problem \cite{BM3}, also in agreement with previously explored empirical results 
\cite{AndrewsLeigh,Corak,WilPic}:
in particular, they exhibited a negative correlation
between economic inequality and social mobility.
With reference to the case without income conservation
they reported a positive correlation between the Gini index and the total income. We regard this as a sign of reliability of the models.
The noise additivity is a perceived drawback though, as it
does not prevent uncontrollably large fluctuations from being compared to class populations, which is unrealistic. 

The goal of this paper is to overcome this limitation by considering instead a multiplicative noise term.
This requires a more subtle procedure
than that proposed in \cite{BCM}.
Attention is then focused on the sign of the correlations between 
income inequality, mobility and total income under different conditions as described in Section \ref{section222} below.

The paper is organized as follows. In the next section, we introduce the structure of the Langevin-type kinetic model. 
Different choices for the construction of the noise term of this structure 
allow to formulate different models. Here, 
in particular, we define two of them: one,
in the first subsection, for which only conservation of the total population holds true,
and 
another, in a second subsection, for which both conservation of total population and total income hold true.
The features of the evolution in time of the solutions of these two models are discussed in Section \ref{section222}.
If the total income $\mu$ is constant in time and not too large, the correlations between the Gini index
and an indicator quantifying social mobility is negative.
The same negativity was obtained in \cite{BCM} in the presence of additive noise, the only difference being the absence of any restriction on the values of $\mu$.
When income conservation does not hold true, the sign of the correlation between the total income and the Gini index
can either be positive or negative, depending on the magnitude of $\mu$, when the noise is multiplicative (the case in the present paper). On the other hand, the correlation is
positive when the noise is additive (the case in \cite{BCM}).
The conclusion summarizes these facts and some directions for future research.

\section{From a deterministic to a Langevin-type kinetic model}
\label{section111}

A simple model describing monetary exchanges between pairs of individuals in a society 
divided into $n$ income classes
can be formulated through a system of
differential equations of the form
\begin{equation}
\frac{{d{x_i}}}{{dt}}(t) = \sum\limits_{h,k = 1}^n {C_{hk}^i{x_h}(t){x_k}(t)} - 
\sum\limits_{h,k = 1}^n {C_{ik}^h{x_i}(t){x_k}(t)} , \qquad \qquad i = 1, ... n .
\label{kin-eq} 
\end{equation}
Here, $x_i(t)$ denotes the fraction of individuals which at time $t$ belong to the $i$-th class
and the constant coefficients $C_{hk}^i \in [0,1]$, such that $\displaystyle \sum_{i=1}^n C_{hk}^i = 1$ for any fixed $h$ and $k$,
express the probability 
that an individual of the $h$-th class will belong to the 
$i$-th class after a direct interaction with an individual of the $k$-th class.
The expression for these coefficients, valid for the case in which 
the average incomes are given by 
\begin{equation}
{r_j} = j \cdot \Delta r ,
\label{rjlinearinj}
\end{equation}
with $\Delta r > 0$, first derived in \cite{B} and then used e.g. also in \cite{BM1, BM2, BMonAMC}, reads when written in compact form:
\begin{eqnarray}
C_{hk}^i  &=&\frac{S}{\Delta r} \, \delta_{hi} \, \Bigg[ \frac{\Delta r}{S} - (1-\delta_{in})(1-\delta_{k1})\:p_{ki} - (1-\delta_{i1})(1-\delta_{kn}) \:p_{ik}
\Bigg] \nonumber \\
&+& \frac{S}{\Delta r} \,
\Bigg[
\delta_{h,i+1} (1-\delta_{kn})\:p_{i+1,k} 
+
\delta_{h,i-1}\:(1-\delta_{k1})\:p_{k,i-1}
\Bigg]
\label{Cihk_eqn}
\end{eqnarray}
with $h,k,i=1,...,n$. Here,
$S << \Delta r$ denotes a unit of money,
${\delta _{hk}}$ denotes the Kronecker's delta, and 
${p_{hk}}$ expresses the probability that in an encounter between an individual of the $h$-th class and one of the $k$-th class, the one
who pays is the former one. We take here
\begin{eqnarray}
{p_{hk}} &=& \frac{1}{{4n}}\min \{ h,k\} \left( {1 - {\delta _{hk}}} \right)\left( {1 - {\delta _{1k}}} \right)\left( {1 - {\delta _{1h}}} \right) 
\left( {1 - {\delta _{nh}}} \right)\left( {1 - {\delta _{nk}}} \right) \nonumber \\
&+& \frac{h}{{2n}}{\delta _{hk}}\left( {1 - {\delta _{1k}}} \right)\left( {1 - {\delta _{nk}}} \right) 
+ \frac{k}{{2n}}{\delta _{nh}}\left( {1 - {\delta _{nk}}} \right)\left( {1 - {\delta _{1k}}} \right) \nonumber \\
&+& \frac{1}{{2n}}{\delta _{1k}}\left( {1 - {\delta _{1h}}} \right)\left( {1 - {\delta _{nh}}} \right) 
+ \frac{1}{{2n}}{\delta _{hn}}{\delta _{k1}} ,
\label{defofphk}
\end{eqnarray}
and extend the
values of ${p_{hk}}$ in (\ref{defofphk}) to allow the indices $h$ and $k$ to go from $0$ to $n+1$
through the definitions ${p_{n + 1,k}} = 0$ for any $k$, and ${p_{k,0}} = 0$ for any $k$.

We emphasize that  the choice of the coefficients (\ref{Cihk_eqn}) is forced, if 
conservation of total income has
to hold true for all $t \ge 0$ once it holds true for $t=0$ (see the proof of Theorem 4.2 in \cite{B}). In contrast, 
there is a certain degree of arbitrariness in the choice of $p_{hk}$. The specific formula (\ref{defofphk}) 
corresponds to a choice made in \cite{BMonAMC}, suggested
by the observation that usually
poor people pay and receive less than rich people. 
Indeed, the ${p_{hk}}$ in (\ref{defofphk}) 
with indices $h$ and $k$ different from $1$ and $n$ are equal to
$\frac{1}{4}\,\frac{\min \{r_h,r_k\}}{r_n}$ \bigg(and those with $h=k$ are equal to $\frac{1}{2}\,\frac{\min \{r_h,r_k\}}{r_n}$\bigg).
The special treatment of the coefficients with indices $h, k = 1$ or $n$ is due to
the impossibility of moving from the first class to a poorer one and from the $n$-th class to a richer one.

A Langevin-type kinetic model \cite{Risken} can now be constructed as a system of stochastic equations of the form
\begin{equation}
{d x_i} = D_i^{(1)}(x)dt + \sum_{j=1}^n D_{ij}^{(2)}(x){\xi _j} \sqrt{\Gamma \:dt} , \qquad \qquad i = 1, ... n ,
\label{Langevin}
\end{equation}
in which the first term on the r.h.s. in Eq. (\ref{Langevin}) represents the \textquotedblleft deterministic\textquotedblright\,
contribution and the second term corresponds to noise.
The interpretation of equation (\ref{Langevin}) is as follows. 
The first term takes into account direct money exchanges,
ruled by norms, and behavioral attitudes which are the same for individuals belonging to the same class.
The second term represents uncertainties randomly occurring, which also affect the 
change in the population distribution.

In the following we take the operator $D_i^{(1)}$ as in (\ref{kin-eq}),
$$
D_i^{(1)}(x) = \sum\limits_{h,k} {C_{hk}^i{x_h}{x_k} - \sum\limits_{h,k} {C_{ik}^h{x_i}{x_k}}} ,
$$
i.e. we take $D_i^{(1)}$ to mimic that component 
of the models in \cite{BM1,BM2} which just describes the direct monetary exchanges 
without taxation and redistribution.
As for the stochastic part, the $\xi_i$ denote $n$ independent Gaussian stochastic variables and
$\Gamma$ denotes the noise amplitude.
The form of the operator
$D_{ij}^{(2)}$
depends on the conservation requirements to which we want the model to obey.

\subsection{Multiplicative noise with conserved total population} 

Enforcing total population conservation, we must get
\begin{equation}
\sum\limits_{i,j}  D_{ij}^{(2)}(x) \xi _j =0 ,
\label{pop-cons}
\end{equation}
for any choice of $\{\xi _j\}$. 
A way to fulfill condition (\ref{pop-cons}) together with a proportionality condition between
the random variations in the class populations
and the population themselves
is to define, starting from the random $\xi_i$, new variations 
$\xi_i'=x_i\xi_i-x_i\sum\limits_{k}x_k \xi_k$, or in matrix form
$$
\xi_i'=\sum\limits_{j}  D_{ij}^{(2)}(x) \xi _j ,
$$
with 
\begin{equation}
D_{ij [pop-const]}^{(2)}(x) = \left\{ \begin{array}{l}
x_i\left(1 - x_i \right),  {\qquad \rm{  if  }}\qquad  i = j\\
- x_i  x_j,   {\qquad  \qquad \rm{  if  }}\qquad  i \ne j .
\end{array} \right. 
\label{firstproposal}
\end{equation}
The formula (\ref{firstproposal}) provides 
an operator $D_{ij [pop-const]}^{(2)}$ which allows to construct,
starting from random variables, 
a multiplicative noise term compatible with the conservation of the total population (\textquotedblleft pop-const\textquotedblright\,).  
Incidentally, we observe that in the following,
as in \cite{BM1,BM2,BM3},
we normalize the total population to $1$.
On the contrary, we emphasize that
allowing a variation of the total income related to stochastic noise amounts to consider for example a society which
also interacts with the 
\textquotedblleft external world\textquotedblright\,: 
capital inflow or outflow
could occur due to 
import-export of goods, incoming-outgoing of tourism, investment and stock trading.

\subsection{Multiplicative noise with conserved total population and income}

Alternatively, we may consider a closed system for which we require
conservation of the total income $\mu=\sum\limits_{i}r_i x_i$\footnote{\ Notice that,
due to the normalization to $1$ of the population, 
the total income coincides with the average income of the population itself.}.
We then point out that from now on we restrict attention on values of $\mu$ satisfying
\begin{equation}
r_1 < \mu < r_n .
\label{constraintsonmu}
\end{equation}
Thanks to the inequalities
(\ref{constraintsonmu}) the possible occurrence can be excluded of situations
in which all individuals belong to the poorest or to the richest income class.
Similar odd cases are not of interest if one wants to deal with realistic situations.
In other words, taking $\mu$ as in (\ref{constraintsonmu}) does not represent a strong assumption.

In addition to (\ref{pop-cons}), a further condition has now to be imposed, i.e.
\begin{equation}
\sum\limits_{i,j}  r_i D_{ij}^{(2)}(x) \xi _j = 0 
\label{incomeconserv}
\end{equation}
for any choice of $\{\xi _j\}$. 
In order to construct a diffusion matrix satisfying both (\ref{pop-cons}) and (\ref{incomeconserv}),
we begin by proving the following proposition.

\bigskip

\noindent {\bf{Proposition 1}}.
Given a vector $x = (x_1, ..., x_n)$ with $x_i > 0$ for all $i$, and $n$ positive constants $r_i$, 
from any vector $\eta_0=(\eta_{01}, ..., \eta_{0n})$ with $|\eta_{0i}| \le 1$  for all $i$,
a new vector $\bar \eta = (\bar \eta_1, ..., \bar \eta_n)$ may be obtained, which satisfies 
the estimates
$$
|{\bar \eta}_i| \le x_i \qquad \hbox{for} \qquad i = 1, ... n
$$
and the two conditions
\begin{equation}
\sum\limits_{i} {\bar \eta}_i=0 \, , \qquad \quad\hbox{and} \quad \qquad \sum\limits_{i} r_i {\bar \eta}_i=0 .
\label{twofirstintegrals}
\end{equation}

\noindent
{\it  {Proof:}} 
We begin by associating to $\eta_0$ a vector 
\begin{equation}
\eta = (\eta_1, ... , \eta_n) = (\frac{x_1 \eta_{0,1}}{C}, ...  , \frac{x_n \eta_{0,n}}{C}) ,
\label{vectoretaproposedfirst}
\end{equation}
where $C \ge 1$ is a constant to be determined in the following. 
We want then to transform the vector $\eta$ to a perturbed vector $\bar\eta = \eta + A \eta$, with components 
\begin{equation}
\bar\eta_i = \eta_i + \sum_{j=1}^n \, a_{ij} \, \eta_j \qquad \hbox{for} \ i = 1, ... n 
\label{etabarequal}
\end{equation}
satisfying
the conservation conditions (\ref{twofirstintegrals}).
Inserting (\ref{etabarequal}) in (\ref{twofirstintegrals}), we 
find (keeping also the arbitrariness of $\eta_{0,i}$ into account) that the conditions (\ref{twofirstintegrals}) become
$$
1+ \sum_{j=1}^n \, a_{ji}  = 0 ,  
\qquad \quad \hbox{and} \quad \qquad 
r_i + \sum_{j=1}^n \, a_{ji} \, r_j  = 0 
$$
for  $i = 1, ... n$. If we choose the matrix $A$ in the set of tridiagonal matrices\footnote{The reason is that with this choice 
the variation of the $i$-th component when passing from $\eta$ to $\bar \eta$ only involves $\eta_{i-1}$, $\eta_{i}$, and $\eta_{i+1}$.},
these conditions read\footnote{Here and henceforth only indexed terms 
with meaningful indices are to be considered present. 
For example, if $i=1$, one has
$\sum_{j=i-1}^{i+1} \, a_{ji} = a_{11} + a_{21}$.}
\begin{equation}
1+ \sum_{j=i-1}^{i+1} \, a_{ji}  = 0 ,  
\qquad \quad \hbox{and} \quad \qquad 
r_i + \sum_{j=i-1}^{i+1} \, a_{ji} \, r_j  = 0 
\label{condsimplytrid}
\end{equation}
for  $i = 1, ... n$.
The formulas (\ref{condsimplytrid}) express $2n$ constraints which the 
$3n-2$ elements $a_{ij}$ of the matrix $A$ have to satisfy\footnote{It is natural to assume $n \ge 3$ here.}.
We then minimize the function of the $3n-2$ variables $a_{ji}$,
$$
f = \sum_{i=1}^n \sum_{j=i-1}^{i+1} \, {a_{ji}}^2  
$$
subject to the $2n$ constraints (\ref{condsimplytrid}).
To this end, we introduce Lagrange multipliers $\lambda_i$ and $\mu_i$ for $i = 1, ... n$, and consider the Lagrangian
\begin{eqnarray*}
L 
=
\sum_{i=1}^n \sum_{j=i-1}^{i+1} \, {a_{ji}}^2 
+ \sum_{i=1}^n \lambda_i \, \Big(1+ \sum_{j=i-1}^{i+1} \, a_{ji}\Big) 
+ \sum_{i=1}^n \mu_i \, \Big(r_i + \sum_{j=i-1}^{i+1} \, a_{ji} \, r_j \Big)  .
\label{newLagrangian}
\end{eqnarray*} 
The search for critical points of $L$ (as a function of the variables $a_{ji}$, $\lambda_i$ and $\mu_i$) 
yields in particular, after straightforward calculations, 
\begin{equation}
a_{ji}  = \frac{N_i \, r_i \, r_j + T_i - R_i \, r_i - R_i \, r_j}{R_i^2 - N_i \, T_i}  ,
\label{lementsaij}
\end{equation}
for $i = 1, ... n, j = i-1,i,i+1$ (the remaining $a_{ji}$ being equal to zero), where
$$
N_1 = 2 , \qquad N_i = 3 \quad \hbox{for} \ i = 2, ... n-1 , \qquad N_n = 2 ,
$$
and$$
R_i = \sum_{k=i-1}^{i+1} r_k \quad \hbox{and} \quad T_i = \sum_{k=i-1}^{i+1}  r_k^2 .
$$
In view of the linearity of $r_j$ in $j$ as
formulated in Eq. (\ref{rjlinearinj}),
it can be easily seen that the matrix $A$ with elements as in (\ref{lementsaij}) takes the form 
$$
A =
\left[
\begin{array}{cccccccccccc}
-1 & -1/3 & 0 & 0 & 0 & ... & ... & 0 & 0 & 0 & 0 & 0 \\ 
0 & -1/3 & -1/3 & 0 & 0 & ... & ... & 0 & 0 & 0 & 0 & 0 \\
0 & -1/3 & -1/3 & -1/3 & 0 & ... & ... & 0 & 0 & 0 & 0 & 0 \\
0 & 0 & -1/3 & -1/3 & -1/3 & ... & ... & 0 & 0 & 0 & 0 & 0 \\
... & ... & ... & ... & ... & ... & ... & ... & ... & ... & ... & ... \\
... & ... & ... & ... & ... & ... & ... & ... & ... & ... & ... & ... \\
0 & 0 & 0 & 0 & 0 & ... & ... & -1/3 & -1/3 & -1/3 & 0 & 0 \\
0 & 0 & 0 & 0 & 0 & ... & ... & 0 & -1/3 & -1/3 & -1/3 & 0 \\
0 & 0 & 0 & 0 & 0 & ... & ... & 0 & 0 & -1/3 & -1/3 & 0 \\
0 & 0 & 0 & 0 & 0 & ... & ... & 0 & 0 & 0 & -1/3 & -1 \\    
\end{array}
\right] .
$$
We observe now that applying the transformation (\ref{etabarequal}) with the matrix $A$ just found, we get
\begin{equation}
\frac{\bar\eta_i}{x_i} =   \frac{\eta_i} {x_i} + \frac{\sum_{j=i-1}^{i+1} a_{ij} \, \eta_j}{x_i} \qquad \hbox{for} \  i = 1, ... n ,
\end{equation}
namely,
\begin{eqnarray*}
& & \frac{\bar\eta_1}{x_1} =  -\frac{1}{3 C} \, \frac{x_2}{x_1} \, \eta_{0,2}  , \\
& & \frac{\bar\eta_2}{x_2} =  \frac{2}{3 C} \, \eta_{0,2} - \frac{1}{3 C} \, \frac{x_3}{x_2} \, \eta_{0,3}  , \\
& & \frac{\bar\eta_i}{x_i} =
\frac{2}{3 C} \, \eta_{0,i} 
- \frac{1}{3 C} \, \frac{x_{i-1}}{x_i} \, \eta_{0,i-1} 
- \frac{1}{3 C} \, \frac{x_{i+1}}{x_i} \, \eta_{0,i+1}  ,
\qquad \hbox{for} \  i = 3, ... n-2 , \\
& & \frac{\bar\eta_{n-1}}{x_{n-1}} =   \frac{2}{3 C} \, \eta_{0,{n-1}} - \frac{1}{3 C} \, \frac{x_{n-2}}{x_{n-1}} \, \eta_{0,{n-2}}  , \\
& & \frac{\bar\eta_n}{x_n} = -\frac{1}{3 C} \, \frac{x_{n-1}}{x_n} \, \eta_{0,n-1}  .
\end{eqnarray*}
For the choice of the constant $C$ appearing here and in (\ref{vectoretaproposedfirst}), we first calculate
\begin{equation}
M_{minus} = \max_{ i = 2, ...n} \, \bigg\{\frac{x_i}{x_{i-1}}\bigg\}  \qquad \hbox{and} \qquad M_{plus} = \max_{i = 1, ...n-1} \, \bigg\{\frac{x_i}{x_{i+1}}\bigg\} ,
\label{Mminusandplus}
\end{equation}
and set
\begin{equation}
\Omega = \max\Big\{1, M_{minus}, M_{plus}\Big\}  .
\label{Omega}
\end{equation}
Then, we fix the constant $C$ in (\ref{vectoretaproposedfirst})
to be equal to $\frac{4}{3} \, \Omega$.
Hence, according to (\ref{vectoretaproposedfirst}), we associate to a randomly chosen vector $\eta_0$ the vector
\begin{equation}
\eta = (\eta_1, ... \eta_n) =  (\frac{3}{4} \, \frac{x_1 \eta_{0,1}}{\Omega}, ... , \frac{3}{4} \,\frac{x_n \eta_{0,n}}{\Omega})  .
\label{costrofeta}
\end{equation}
Now, applying to $\eta$ the transformation (\ref{etabarequal}) with the ${a_{ij}}'s$ as in (\ref{lementsaij}), we get
\begin{eqnarray*}
& & \frac{\bar\eta_1}{x_1} =  \frac{3}{4} \, \bigg (-\frac{1}{3} \, \frac{x_2}{x_1} \, \frac{1}{\Omega} \, \eta_{0,2}   \bigg ) , \\
& & \frac{\bar\eta_2}{x_2} =  \frac{3}{4} \, \bigg (\frac{2}{3} \, \frac{1}{\Omega} \, \eta_{0,2} - \frac{1}{3} \, \frac{x_3}{x_2} \, \frac{1}{\Omega} \, \eta_{0,3}  \bigg ) , \\
& & \frac{\bar\eta_i}{x_i} =
\frac{3}{4} \, \bigg (\frac{2}{3} \, \frac{1}{\Omega} \, \eta_{0,i} 
- \frac{1}{3} \, \frac{x_{i-1}}{x_i} \, \frac{1}{\Omega} \, \eta_{0,i-1} 
- \frac{1}{3} \, \frac{x_{i+1}}{x_i} \, \frac{1}{\Omega} \, \eta_{0,i+1}  \bigg ) , 
\qquad \hbox{for} \  i = 3, ... n-2  , \\
& & \frac{\bar\eta_{n-1}}{x_{n-1}} =   \frac{3}{4} \, \bigg (\frac{2}{3} \, \frac{1}{\Omega} \, \eta_{0,{n-1}} - \frac{1}{3} \, \frac{x_{n-2}}{x_{n-1}} \, \frac{1}{\Omega} \, \eta_{0,{n-2}}  \bigg ) , \\
& & \frac{\bar\eta_n}{x_n} = \frac{3}{4} \, \bigg (-\frac{1}{3} \, \frac{x_{n-1}}{x_n} \, \frac{1}{\Omega} \, \eta_{0,n-1} \bigg ) ,
\end{eqnarray*}
which in turn implies
\begin{eqnarray*}
& & \bigg |\frac{\bar\eta_1}{x_1}\bigg | \le  \frac{1}{4}  \, |\eta_{0,2}|  \le 1 , \\
& & \bigg |\frac{\bar\eta_2}{x_2}\bigg | \le  \frac{2}{4} \, |\eta_{0,2}| + \frac{1}{4}  \, |\eta_{0,3}|  \le 1 , \\
& & \bigg |\frac{\bar\eta_i}{x_i}\bigg | \le
\frac{2}{4} \, |\eta_{0,i}| 
+ \frac{1}{4} \, |\eta_{0,i-1}| 
+ \frac{1}{4} \,  |\eta_{0,i+1}|  \le 1 ,
\qquad \hbox{for} \  i = 3, ... n-2 , \\
& & \bigg|\frac{\bar\eta_{n-1}}{x_{n-1}}\bigg | \le   \frac{2}{4} \, |\eta_{0,{n-1}}| + \frac{1}{4} \, |\eta_{0,{n-2}}|  \le 1 , \\
& & \bigg |\frac{\bar\eta_n}{x_n}\bigg | \le \frac{1}{4} \, |\eta_{0,n-1}|  \le 1.
\end{eqnarray*}
In conclusion, the vector $\bar\eta$
satisfies the conservation conditions given in Eq. (\ref{twofirstintegrals}) 
as well as the estimates $|\bar\eta_i| \le x_i$ for $ i = 1, ... n$.
$\square$

\bigskip

In order to construct from the 
stochastic variable $\xi$ a multiplicative noise term 
satisfying conservation of population and income, one can discretize time and repeatedly iterate,
as illustrated below,
the procedure of Proposition $1$.
We emphasize that a warning as discussed in the next lines is in order here. 

At each step, say at each time $t_k$ with $k = 0, 1, 2, ...$, a vector
$\xi$
is picked
whose components $\xi_i$ for $i = 1, ... n$ 
are Gaussian random numbers ranging from $-1$ to $1$.
Here, $\xi$ plays the role of $\eta_0$ in Proposition $1$.
The vector $x = (x_1, ..., x_n)$ of Proposition $1$
is given at the beginning of the process, i.e. at time $t_0$,
by a stationary distribution $x_{eq}$ (reached in the long run) of the \textquotedblleft deterministic\textquotedblright\, system (\ref{kin-eq}),   
whereas at subsequent steps, i.e. at time $t_k$ with $k = 1, 2, ...$, it is given by the solution $x(t_k)$ of the system (\ref{Langevin}), 
or of the system (\ref{kin-eq}), according to the criterion described next. 
There are two possibilities: either $x_i > 0$ for all $i = 1, ... n$ or 
there exists at least an index $i^*$ such that $x_{i^*}$ vanishes.\footnote{In fact it is highly improbable that the second alternative occurs. Nevertheless, we
take it too into consideration.}
A control loop in the algorithm checks 
which of the two possibilities holds true.
Accordingly, the procedure to be applied is as follows.

\begin{enumerate}
\item If at time $t_k$ it is $x_i > 0$ for all $i = 1, ... n$, one calculates $\Omega$ according to 
Eq.s (\ref{Mminusandplus}) and (\ref{Omega}) and then defines,
by applying the formula (\ref{costrofeta}) with this value of $\Omega$, an \textquotedblleft intermediate\textquotedblright\, vector $\eta$. Then,
one applies to $\eta$ the transformation (\ref{etabarequal}) with the ${a_{ij}}'s$ as in (\ref{lementsaij}).
In this way one obtains,
as Proposition $1$ shows, a vector whose components are proportional to the classes populations
and which, when inserted in the equation (\ref{Langevin}), guarantees both population and total income conservation (\textquotedblleft pop-inc-const\textquotedblright\,).
This vector can be denoted by
\begin{equation}
D_{[pop-inc-const]}^{(2)}(x) \xi .
\label{secodproposal}
\end{equation}
Numerical solutions of (\ref{Langevin}) can be found 
by calculating (\ref{secodproposal}), inserting the noise term
$$
D_{ij[pop-inc-const]}^{(2)}(x) {\xi_j}  \sqrt{\Gamma \:dt}
$$ 
into the equation (\ref{Langevin}) and
getting the corresponding solution $x(t_{k+1})$.
If $x_i(t_{k+1})>0$ for all $i = 1, ... n$ and all $k \in \bf{N}$, one repeats all this over and over again.
\item If for some some integer $k$ and some index $i^*$, the component $x_{i^*}(t_k)$ vanishes, 
i.e. denoting $t_k=t^*$ one has $x_{i^*}(t^*)=0$,
then one lets only the system
(\ref{kin-eq}) evolve, without adding any noise up to when
$x_i > 0$ for all $i = 1, ... n$. From then on,
the algorithm described in $1$ has to be applied again.
To give an insight as to why the re-establishment of the situation with all $x_i > 0$ is
to be 
expected, we argue as follows.

First of all, we want to exclude the cases (both of which are equilibria for the system (\ref{kin-eq})),
for which all individuals belong to the poorest class or to the richest class.
Since the value of the total income with which the former case is compatible is $\mu = r_1$
whereas for the latter case it is $\mu = r_n$,
the assumption (\ref{constraintsonmu}) guarantees that these cases cannot occur,
thereby assuring \textquotedblleft moderate income\textquotedblright\, remit.

We then observe that exploiting the fact that $x_{i^*}(t^*) = 0$ one gets from (\ref{kin-eq}),
\begin{equation}
\frac{{d{x_{i^*}}}}{{dt}}(t^*) = \sum\limits_{h\ne {i^*}} \sum\limits_{k \ne {i^*}} {C_{hk}^{i^*}{x_h}(t^*){x_k}(t^*)} \ge 0 .
\end{equation}
It is of course possible that other $x_i$ in addition to $x_{i^*}$ vanish at time $t^*$. Then, let
$m$ be the smallest positive integer such that
\begin{equation}
x_{i^*-m}(t^*) \ne 0 \qquad \hbox{or} \qquad x_{i^*+m}(t^*) \ne 0
\label{differentfromzero}
\end{equation}
holds true. Such a number certainly exists. Assume, without loss of generality, the second of the two inequalities (\ref{differentfromzero}) to hold true.
The other case can be handled similarly.
Now, either $i^*+m < n$ or $i^*+m =n$ holds true.
\begin{itemize}
\item If $i^*+m < n$, observing that $C_{ii}^{i-1} > 0$ (as also $C_{ii}^{i+1} > 0$) provided $1 < i < n$, we conclude that $C_{i^*+m\, ,\, i^*+m}^{i^*+m-1} > 0$ and hence
$$
{\frac{d}{dt}} {x_{i^*+m-1}} (t^*) \ge {C_{i^*+m\, ,\, i^*+m}^{i^*+m-1} \, {x^2_{i^*+m}}(t^*)}> 0 .
$$
Consequently, $x_{i^*+m-1} (t^*+1) > 0$. Iterating the procedure $m$ times, one obtains
$$
x_{i^*} (t^*+m) > 0 .
$$
\item If $i^*+m =n$, we know, in view of (\ref{constraintsonmu}), that there exists a positive integer $p$, satisfying $1 \le i^*-p$, such that $x_{i^*-p}(t^*) \ne 0$.
If $i^*-p > 1$, then, similarly as above, one notices that
${\frac{d}{dt}} {x_{i^*-p+1}} (t^*) \ge {C_{i^*- p\, ,\, i^*- p}^{i^*- p +1} \, {x^2_{i^*- p}}(t^*)}> 0$, from which $x_{i^*-p+1} (t^*+1) > 0$ and then $x_{i^*} (t^*+p) > 0$
follows.
If $i^*-p = 1$, then one may exploit the fact that $C_{1n}^{2} > 0$ and ${\frac{d}{dt}} {x_{i^*-p+1}} (t^*) \ge {C_{1n}^{2} \, {x_1(t^*)}{x_n(t^*)}}> 0$ to be 
reconduced to the case just dealt with.
\end{itemize}
By repeating, if necessary, the procedure here illustrated, one ends up with $x_i(t_{k+q})>0$ for all $i = 1, ... n$, for some $q \in \bf{N}$.
\end{enumerate}

\section{Simulation results}
\label{section222}

To investigate the stochastic processes of the two models designed in Section \ref{section111}, we 
numerically solved the equations (\ref{Langevin}) and took the average of various quantities over a large number of stochastic realizations.
Of course, 
no equilibria have to be expected in the present case.
To draw some conclusions, we need to recall the definition - more precisely, a variant of it, suitable for the present case -
of an indicator of social mobility introduced in \cite{BM3}. 
This indicator,
which expresses the collective probability of class advancement of all classes from the $2$-th to the $(n - 1)$-th one,
is given by
$$
M =  \frac{1}{(1 - {x}_{1} - {x}_{n})} \, \sum_{i=2}^{n-1} \,\sum_{k=1}^{n} \, \frac{S}{{{(r_{i + 1}} - {r_i})}} \, {{p_{k,i}}{{x}_k}{{x}_i}}.
$$
We calculated the value of $M$ in a succession of equally spaced instants $\{t_j\}$
along the evolution in time of several solutions of Eq. (\ref{Langevin}).
As well, in correspondence to the same instants, we calculated the Gini index $G$\footnote{This coefficient
was introduced by the italian statistician Corrado Gini a century ago. It takes values in $[0,1]$ and it is defined as a ratio, having the numerator given by the area
between the Lorenz curve of a distribution and the uniform distribution line,
and the denominator given by the area of the region under the uniform distribution line.}.

\medskip

\begin{figure}[htbp]
\begin{center}
\includegraphics[width=6cm,height=3cm] {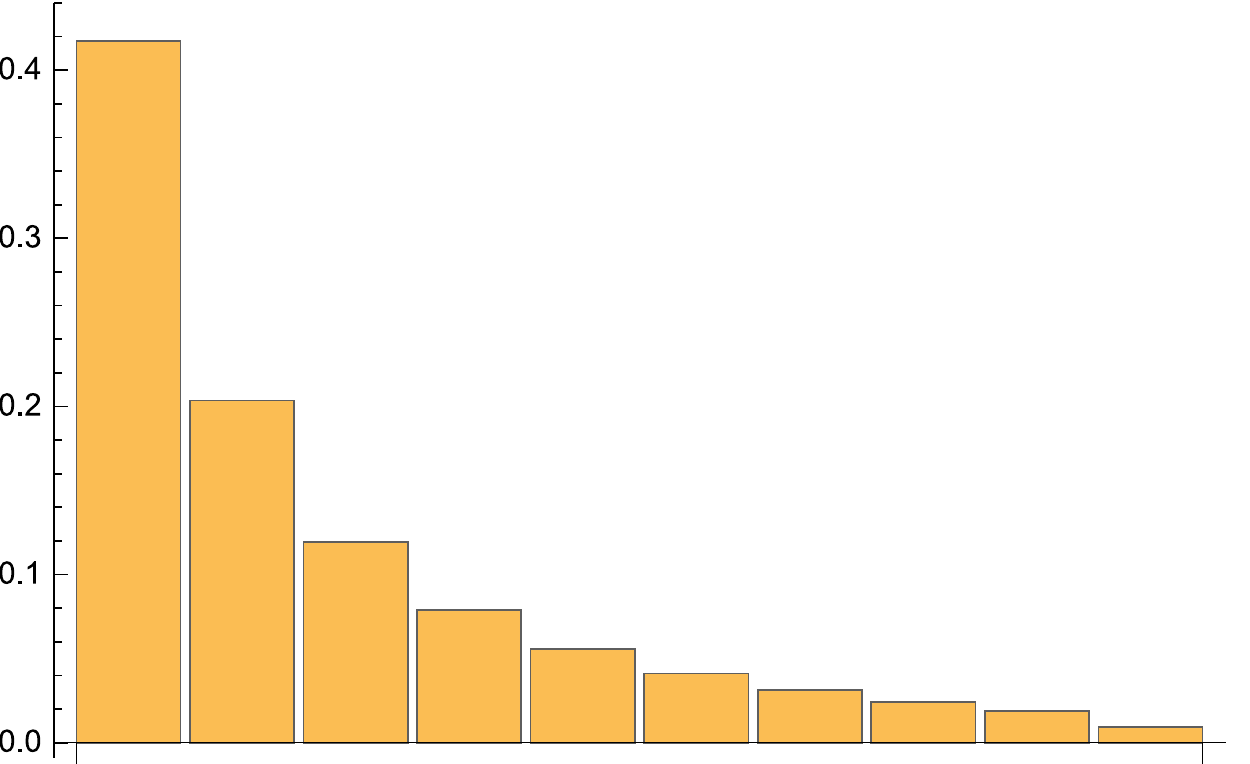}
\end{center}
\caption{The asymptotic stationary solution of the \textquotedblleft deterministic\textquotedblright\, system with constant total income $\mu = 27$.}
\label{fig:asymptoticdeterministicsol}
\end{figure}

A significant finding concerns the sign of the correlation between $G$ and $M$, namely between economic inequality and social mobility.
For values of the total income 
$\mu$ which are not too large when total income is conserved,
and which are neither too large nor too small when total income is not conserved, the statistical value of the sign of the correlation between $G$ and $M$ turns out to be negative. 
The values of $\mu$ under consideration\footnote{For example, if we take $n=10$ and fix
the values of $r_i$ for $i=1,...,n$ to be linearly growing from $r_1=10$ to $r_{10}=100$,
values of $\mu \le 30$ in the conservative case and $\mu\in [24,30]$ in the non conservative case meet this criterion.} are reasonable in a realistic perspective (see e.g. \cite{PRC}) because they are compatible with a distribution of individuals 
in which most of the population belongs to the low-middle classes.
The negativity of the correlation which we get is in agreement with a great deal of empirical data \cite{AndrewsLeigh,Corak} 
and provides evidence of some robustness
against random perturbations of the corresponding property established for systems without noise in \cite{BM3}.
A few samples of correlations $R_{G M}$ (Gini and mobility index) are given in Table \ref{tab:tableRGMconserved} for the case with constant total income and in Table \ref{tab:tableRGMnonconserved} for the case with varying total income. For our simulations, we considered
the difference $\Delta r$ between class average incomes 
equal to $10$ and the noise amplitude $\Gamma$ equal to $0.001$.
The correlations were obtained as averages of $50$ realizations, each over $5000$ integration steps. 
In these samples, three initial conditions - the same in the Tables \ref{tab:tableRGMconserved} and \ref{tab:tableRGMnonconserved} - 
compatible with values of the total income equal to $24.5$, $27$ and $29.5$ respectively, are considered
(Figure \ref{fig:asymptoticdeterministicsol} displays
the initial condition corresponding to the asymptotic stationary distribution for the
system without noise (\ref{kin-eq}) with $\mu = 27$).
And for each of these initial conditions, three different average results are reported. 

We also stress here that the distributions we get after the $5000$ integration steps
remain in fact quite \textquotedblleft close\textquotedblright\, to the distributions from which they evolve,
which are equilibria if noise is absent. We measure the  \textquotedblleft closeness\textquotedblright\, by calculating in correspondence to each realization
the difference between the average (of the $5000$ values attained during evolution) $\hat x_i$ of each component of the distribution
and the corresponding initial value $x_i(0)$; in addition, we calculate
the standard deviation $\sigma_{x_i}$ of each component $x_i$. We find that 
the differences $\hat x_i - x_i(0)$ take values whose order of magnitude typically
are between $10^{-5}$ and $10^{-7}$,
the $\sigma_{x_i}$ 
take values whose order of magnitude typically oscillate between $10^{-4}$ and $10^{-6}$, whereas the 
values of the relative standard variations $\sigma_{x_i}/x_i$ typically are of the order of $10^{-4}$.

\begin{table}[htbp]
\begin{center}
\begin{tabular}{|c|c|c|c|}
\hline
$\mu$ & $R_{G M}$ & $R_{G M}$ & $R_{G M}$ \\
\hline\hline
24.5 & - 0.980 $\pm$ 0.002 & - 0.984 $\pm$ 0.001 & - 0.983 $\pm$ 0.002 \\
\hline
27.0 & - 0.967 $\pm$ 0.003 & - 0.970 $\pm$ 0.003 & - 0.968 $\pm$ 0.003 \\
\hline
29.5 & - 0.913 $\pm$ 0.007 & - 0.923 $\pm$ 0.008 & - 0.920 $\pm$ 0.007 \\
\hline
\end{tabular}
\end{center}
\caption{Correlations $R_{GM}$ (Gini and mobility index) 
computed in nine cases in which total income $\mu$ is conserved, 
with noise amplitude $\Gamma=0.001$. Averages of 50 realizations, each of 5000 integration steps.}
\label{tab:tableRGMconserved}
\end{table}

\begin{table}[htbp]
\begin{center}
\begin{tabular}{|c|c|c|c|}
\hline
$\mu(0)$ & $R_{G M}$ & $R_{G M}$ & $R_{G M}$ \\
\hline\hline
24.5 & - 0.150 $\pm$ 0.061 & - 0.204 $\pm$ 0.056 & - 0.220 $\pm$ 0.062 \\
\hline
27.0 & - 0.276 $\pm$ 0.064 & - 0.475 $\pm$ 0.051 & - 0.450 $\pm$ 0.052 \\
\hline
29.5 & - 0.610 $\pm$ 0.044 & - 0.611 $\pm$ 0.034 & - 0.605 $\pm$ 0.047 \\
\hline
\hline
\hline
$\mu(0)$ & $R_{G \mu}$ & $R_{G \mu}$ & $R_{G \mu}$ \\
\hline\hline
24.5 & \, 0.096 $\pm$ 0.061 & \, 0.043 $\pm$ 0.059 & \, 0.045 $\pm$ 0.063 \\
\hline
27.0 & - 0.068 $\pm$ 0.067 & - 0.271 $\pm$ 0.059 & - 0.239 $\pm$ 0.058 \\
\hline
29.5 & - 0.465 $\pm$ 0.052 & - 0.443 $\pm$ 0.043 & - 0.466 $\pm$ 0.054 \\
\hline\end{tabular}
\end{center}
\caption{Correlations $R_{GM}$ (Gini and mobility index) 
and $R_{G\mu}$ (Gini index and total income)
computed in nine cases in which total income $\mu$ is not conserved, 
with noise amplitude $\Gamma=0.001$. Averages of 50 realizations, each of 5000 integration steps.}
\label{tab:tableRGMnonconserved}
\end{table}

\smallskip

\begin{figure}
\hfill
\subfigure[$G$ vs $t$]{\includegraphics[width=4cm]{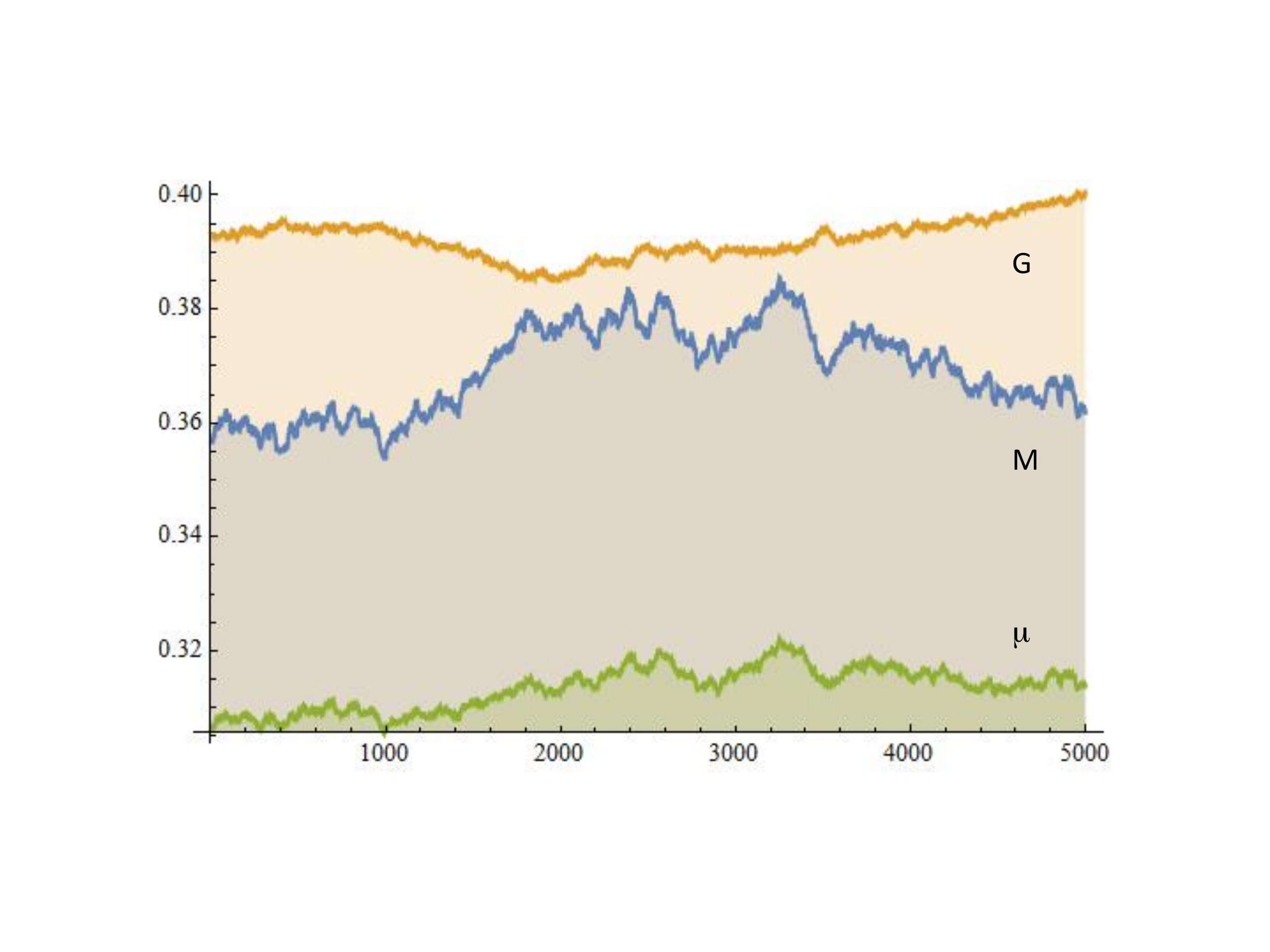}}
\hfill
\subfigure[$M$ vs $t$]{\includegraphics[width=4cm]{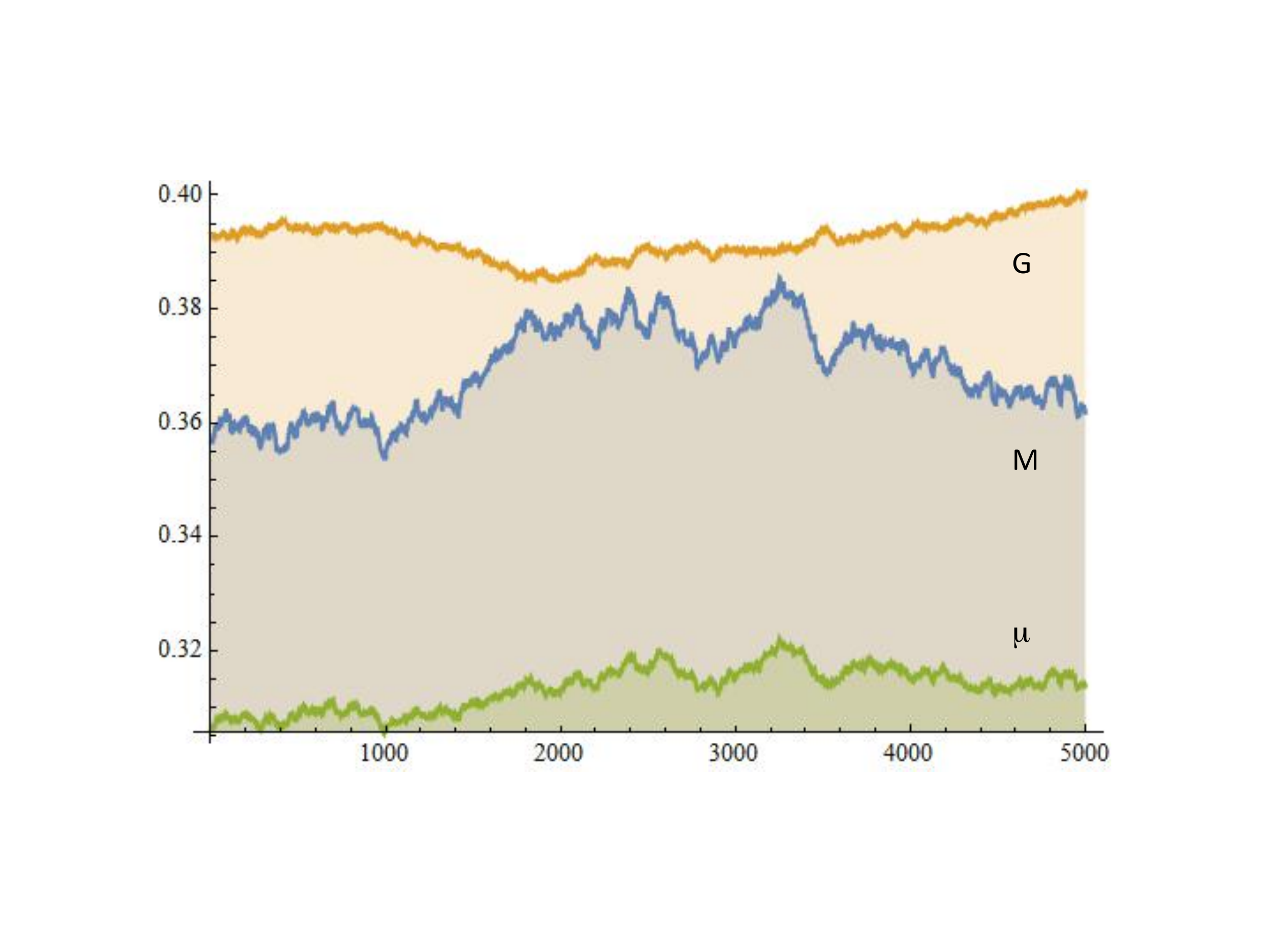}}
\hfill
\subfigure[$\mu$ vs $t$]{\includegraphics[width=4cm]{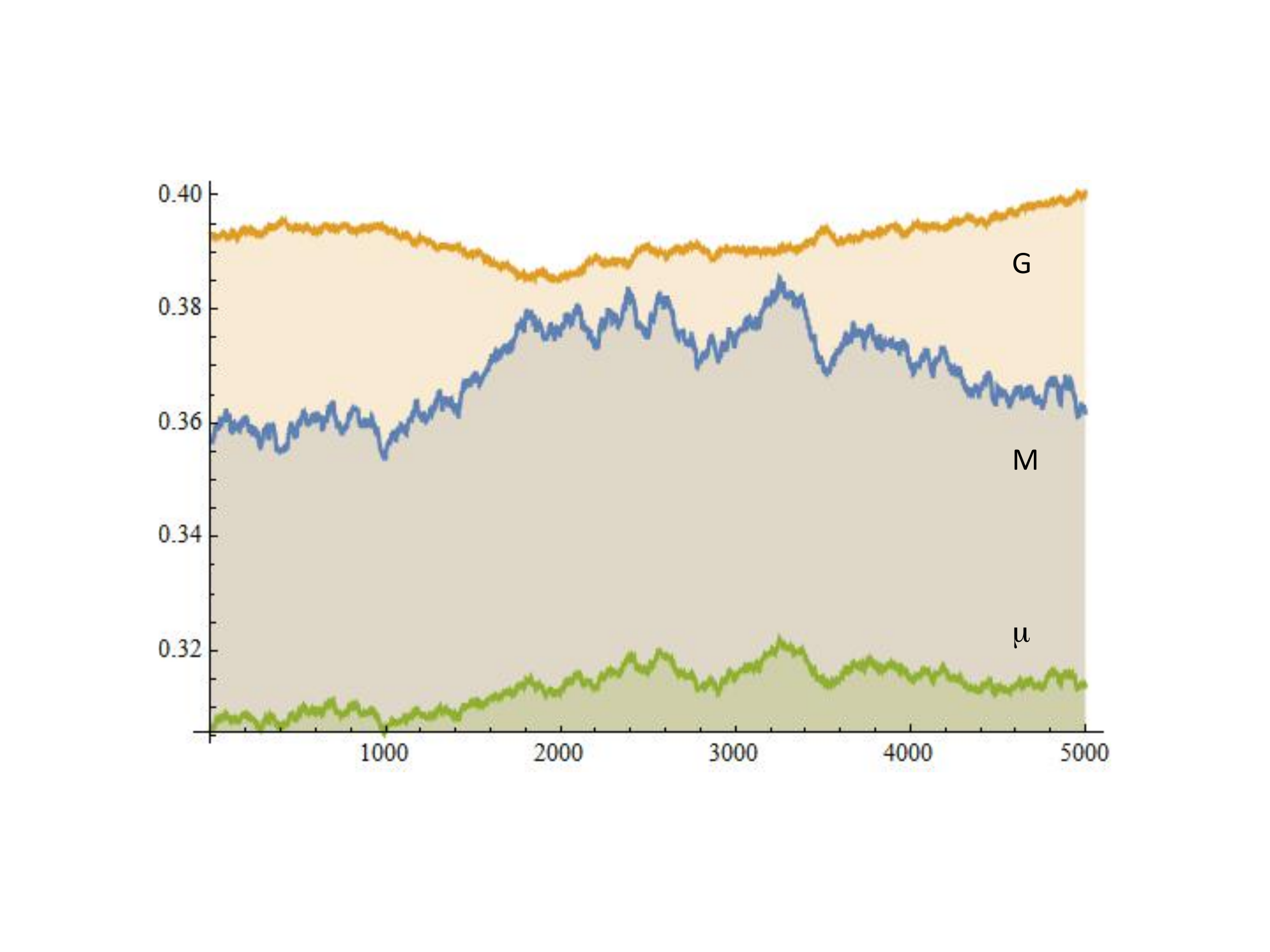}}
\caption{
Samples of time series of $G$, $M$ and $\mu$ in three cases with
$\mu(0)$ equal to $24.5$, $27$ and $29.5$ respectively.
The values of $M$ are here multiplied by $800$ and those of $\mu$ are divided by $80$ so as to obtain a clearer comparison.
In particular, a negative correlation between $G$ and $M$, as well as a positive correlation between $M$ and $\mu$ are clearly visible.}
\label{fig:graphsofcorrelations}
\end{figure}

A further issue which
one can explore in the non conservative case 
is
the correlation $R_{G \mu}$ between the Gini index and the total income.
A difference comes out in this respect, depending on whether
the noise is additive or multiplicative: whereas the value of
$R_{G \mu}$ provided by the numerical simulations is positive in the first case,
it turns out to be sometimes negative and sometimes positive in the second one,
depending on the value of the initial total income $\mu$.
An intuitive argument for a possible explanation 
of the positive sign in the additive case is as follows:
in the presence of additive noise the variations in the rich classes
are typically much larger (with respect to those in the low and middle classes) than when the noise is multiplicative.
This causes larger variations in the total income.
Since
increases of $\mu$ mainly affect the richer classes, this
brings about an increase of inequality, i.e. of $G$.
Yet, we do not have an explanation for the behavior of the correlation $R_{G \mu}$ in the multiplicative case.
It has also to be noticed that the values reported in the Table \ref{tab:tableRGMnonconserved} 
evidentiate
a great variability (and possibly, even no meaningfulness) of $R_{G \mu}$, 
when the total income is not fixed.
We notice however that a strong positive correlation $R_{M \mu}$
between mobility and total income
comes out of the realizations. A few samples of that are reported in the Table \ref{tab:tableRMmu}.
Also, from the three panels in Figure \ref{fig:graphsofcorrelations} displaying time series of $G$, $M$ and $\mu$
the negativity of the correlation between $G$ and $M$
and
the positivity of the correlation between $M$ and $\mu$
is clearly visible.

\begin{table}[htbp]
\begin{center}
\begin{tabular}{|c|c|c|c|c|c|}
\hline
$\mu(0)$ & 22.0 & 24.5 & 27.0 & 29.5 & 32.0\\
\hline\hline
$R_{M \mu}$ & 0.951 $\pm$ 0.007 & 0.950 $\pm$ 0.006 & 0.960 $\pm$ 0.006 & 0.972 $\pm$ 0.005 & 0.981 $\pm$ 0.004 \\
\hline
\end{tabular}
\end{center}
\caption{Correlations $R_{M\mu}$ (mobility index and total income) 
computed in five cases with different values of the initial total income $\mu$. Again,
noise amplitude $\Gamma$ is equal to $0.001$ and averages are taken out of 50 realizations, each of 5000 integration steps.}
\label{tab:tableRMmu}
\end{table}

\begin{figure}
\hfill
\subfigure[$R_{MG}$ versus $G$]{\includegraphics[width=7cm]{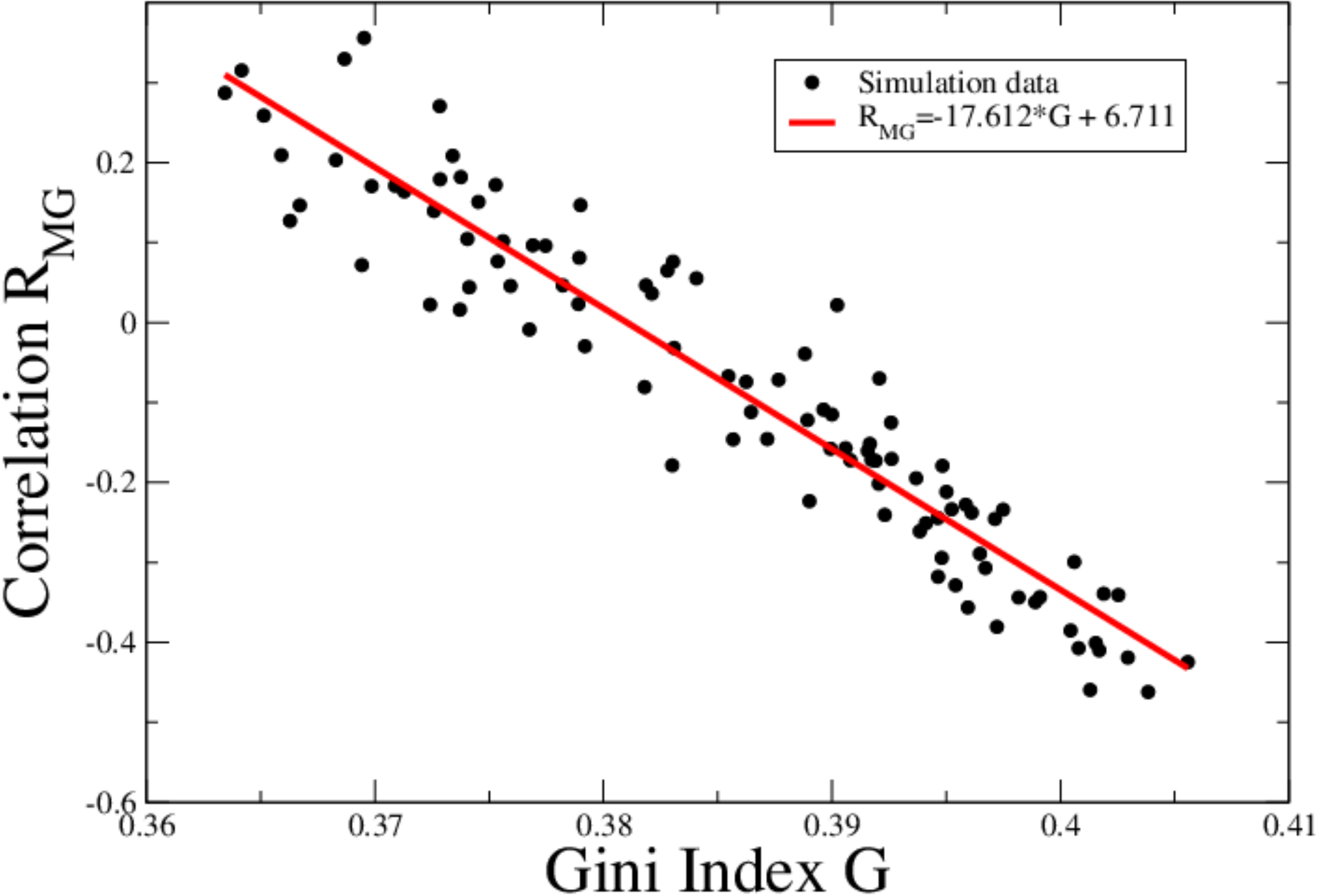}}
\hfill
\subfigure[$R_{\mu G}$ versus $G$]{\includegraphics[width=7cm]{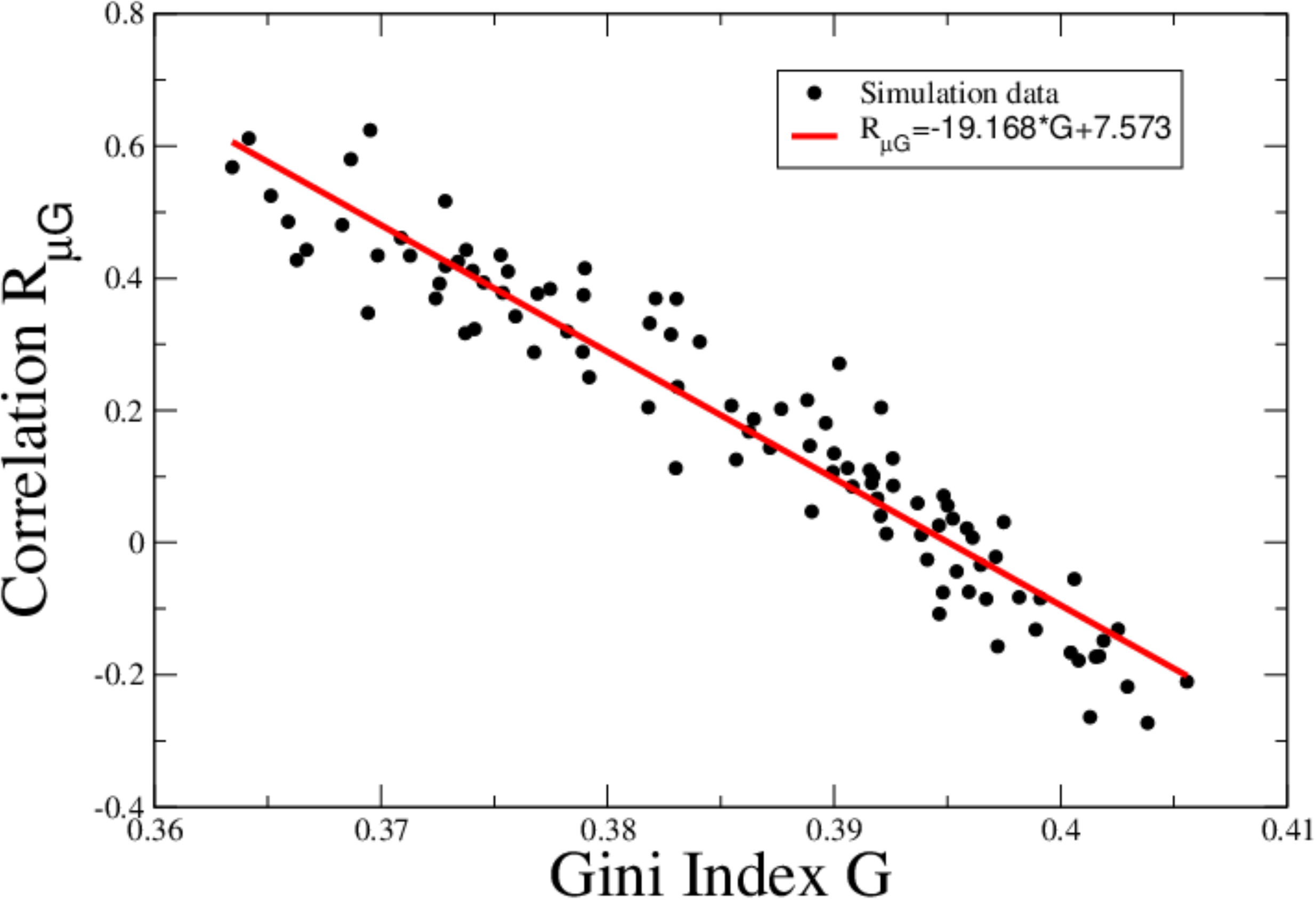}}
\hfill
\caption{Part (a): Correlation between the total income $\mu$ and the Gini index $G$. Part (b): Correlation between the mobility $M$ and the Gini index $G$.}
\label{fig3}
\end{figure}

\begin{figure}[htbp]
\begin{center}
\includegraphics[width=10cm,height=8cm] {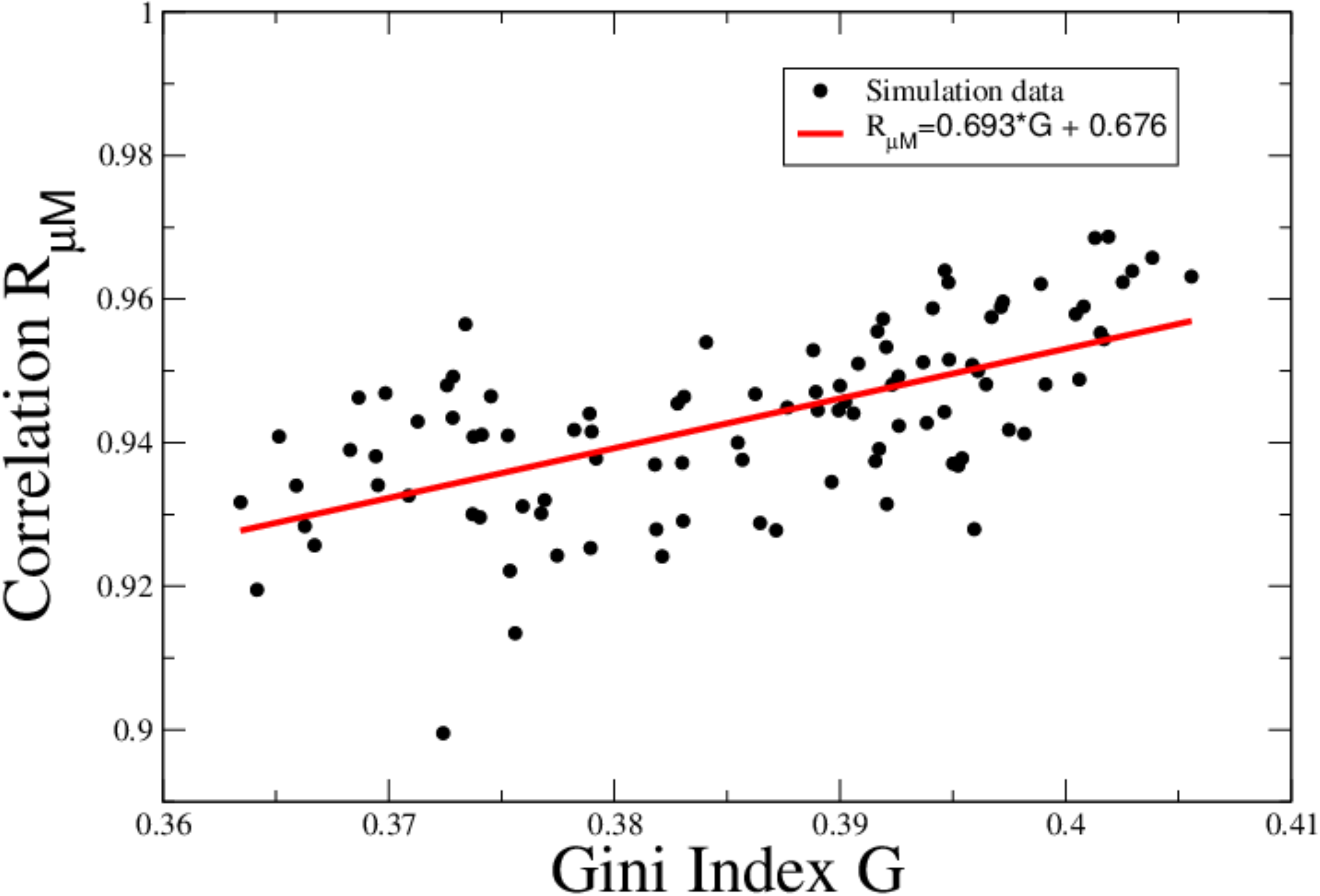} 
\end{center}
\caption{Correlation between the total income $\mu$ and the mobility $M$.}
\label{figRmuM}
\end{figure}

As can be seen in the Tables \ref{tab:tableRGMnonconserved} and \ref{tab:tableRMmu} the correlations between $G$, $M$ and $\mu$ depend on the value of $\mu$. Since the values of $\mu$ and $G$ at equilibrium are mutually related, the correlations depend on $G$. In order to further check this dependence, we ran simulations over 100 cycles varying $\mu$, the results of which are shown in Figure \ref{fig3}. $\mu$ is varied approximately between $21 < \mu <28$, corresponding for $G$ to $0.36 < G < 0.41$. Each simulation consists of 50 stochastic realizations, each over 5000 steps and starting from the same equilibrium configuration; the solid circles in the plot represent the simulation data.

Figure \ref{fig3}a shows that the $M$-$G$ correlation is positive in the interval $0.36 < G < 0.38$; for $G>0.38$, the aforementioned correlation becomes negative. Therefore, according to our model, the \textquotedblleft Great Gatsby law\textquotedblright, which states that the correlation between inequality and economic mobility is negative, strictly holds for $G>0.38$. This is actually a range representing the pre-taxation values of $G$ that includes most industrialized countries.

Figure \ref{fig3}b shows that the $\mu$-$G$ correlation is positive in the interval $0.36 < G < 0.395$ but gets negative thereafter. We thus identify a window of values for $G$ for which the influx of wealth to the system contributes in decreasing inequality.

Finally, Figure \ref{figRmuM} shows that the total income $\mu$ and mobility $M$ always have a strong positive correlation which shows a slow increase with increasing $G$. This could be understood from an established thermodynamic allusion: in a canonical ensemble, for any reasonable definition of mobility, we expect a strong positive correlation between mobility and temperature; and in turn temperature variations will be strongly correlated with the variations in the free energy (corresponding to income in our case). 

\section{Conclusion} 

In this article, we proposed two different models to analyze the time evolution of income distribution
resulting from multiple economic exchanges, in the presence of a multiplicative noise (abiding the Ito formulation). The presence of noise causes a continuous dynamical adjustment of the income distribution, while still staying reasonably close 
to the large time steady state limit that
it would have reached in the absence of noise.
Ensemble averaging over a large set of stochastic realizations, we observed the emergence of correlations between the Gini inequality index $G$ and a suitably defined mobility index $M$. The respective correlations between mobility $M$ with the Gini index $G$ and that between total income $\mu$ with the Gini index $G$, for the time varying case as we consider here, depict association between these quantities. Both the mobility $M$ and the total income $\mu$ show steady decrease with increasing $G$, a reflection of the fact that an increasing inequality contributes to decreased social mobility (Figure \ref{fig3}). On the other hand, an increasing inequality (reflected by an increasing value of $G$) portrays the strength of interaction between the social mobility and total income which then shows a marginal steady increase (Figure \ref{figRmuM}). Some relevant comparisons with results from an equivalent additive noise case are also drawn.

Probably, a more realistic model should involve a weighted combination of both additive and multiplicative stochastic perturbation. Indeed, certain events act as additive noise, whereas others are more properly represented by multiplicative noise. While economics modeling is replete with examples of application of additive noise \cite{Chattopadhyay2011,Dechant2017}, implementation of multiplicative noise is also not unknown \cite{ Mantegna1995, Laloux1999, Plerou1999}. 

Correlated noise spectra, like Ornstein-Uhlenbeck, as used in other branches of material science \cite{Chattopadhyay2016}, could be considered as well, which is one of our ongoing research projects. More complicated noise structures, resembling power-law scaling have found popular applications in cognition science \cite{Wagenmakers2004}, another possibility for future investigation.

A further extension of the models developed in \cite{BCM} and here could involve studying the impact of the coefficients ${C^{i}_{hk}}$ themselves changing with the income distribution. Finally, it would be of great interest to investigate the  dependence of the entire dynamical process, both on the amplitude as also on the nature of the noise distribution, as alluded to in some of the earlier references in other fields.

\end{document}